\documentclass[fleqn,usenatbib]{mnras}

\usepackage{newtxtext,newtxmath}

\usepackage[T1]{fontenc}
\usepackage{ae,aecompl} 
\usepackage{afterpage}
\usepackage{xcolor}

\usepackage{graphicx}	
\usepackage{amsmath}	

\newcommand{\be}{\begin{equation}}
\newcommand{\ee}{\end{equation}}
\newcommand{\ms}{_{\odot}}
\newcommand{\bh}{_{\bullet}}

\title[GW-EMRIs: mass-segregation and binary disruptions by MBHs]{Extreme mass-ratio gravitational-wave sources:\\ Mass segregation and post binary tidal-disruption captures}

\author[Raveh \& Perets]{
Yael Raveh\thanks{E-mail: yael.raveh@campus.technion.ac.il}
\& Hagai B. Perets
\\
Faculty of Physics, Technion -- Israel Institute of Technology, Haifa, 3200003, Israel
}

\date{Accepted XXX. Received YYY; in original form ZZZ}

\pubyear{2020}

\begin{document}
\label{firstpage}
\pagerange{\pageref{firstpage}--\pageref{lastpage}}
\maketitle

\begin{abstract}
The gravitational-wave (GW) inspirals of stellar-mass compact objects onto a supermassive black hole (MBH), are some of the most promising GW sources detectable by next-generation space-born GW-detectors. The rates and characteristics of such extreme mass ratio inspirals (EMRIs) sources are highly uncertain. They are determined by the dynamics of stars near MBHs, and the rate at which compacts objects are driven to the close proximity of the MBH. Here we consider weakly and strongly mass-segregated nuclear clusters, and the evolution of stars captured into highly eccentric orbits following binary disruptions by the MBH. We make use of a Monte-Carlo approach to model the diffusion of both captured objects, and compact-objects brought through two-body relaxation processes. We calculate the rates of GW-inspirals resulting from relaxation-driven objects, and characterize EMRIs properties. We correct previous studies and show that relaxation-driven sources produce GW-sources with lower-eccentricity than previously found, and provide the detailed EMRI eccentricity distribution in the weak and strong mass-segregation regimes. We also show that binary-disruption captured-stars could introduce low-eccentricity GW-sources of stellar black-hole EMRIs in mass-segregated clusters. The eccentricities of the GW-sources from the capture channel, however, are strongly affected by relaxation processes, and are significantly higher than previously suggested. We find that both the rate and eccentricity distribution of EMRIs could probe the dynamics near MBHs, and the contribution of captured stars, characterize the mass-function of stellar compact objects, and verify whether weak or strong mass-segregation processes take place near MBHs.    
\end{abstract}

\begin{keywords}
methods: numerical -- gravitational waves -- black hole physics -- Galaxy: kinematics and dynamics -- Galaxy: center -- binaries: general
\end{keywords}

\section{Introduction}
The era of low-frequency gravitational wave (GW) astronomy is likely to begin in the coming few decades with the planned launch of the Laser Interferometer Space Antenna (LISA) GW detector. Key targets for LISA are extreme mass ratio inspirals (EMRIs), which have been the focus of many past studies \citep{2003astro.ph..7084L,0eccLISA,HA05,HA06seg,Danor,2016ApJ...820..129B,2017PhRvD..95j3012B}. The large mass ratio associated with EMRIs, makes them effectively behave as test masses in the interaction with the massive black hole (MBH). Thus, EMRIs are expected to provide an unprecedented probe of the properties of MBHs \citep{2010PhRvD..81j4014G}, and the dynamical interactions in their environment  and  shed light on relaxation processes occurring in galactic nuclei and the structures of nuclear stellar clusters (NSCs).

Previous EMRI rate estimates have focused on the capture of compact objects (COs) by emission of gravitational radiation during close passes.  
In the canonical model, the stellar objects are pushed onto sufficiently bound orbits due to series of random two-body scatterings that bring them close enough to the MBH, where dissipation by GW emission is significant. When the orbits become detectable with LISA, they have significant eccentricities of typically $e \sim 0.5-0.9$ \citep{HA05}. 

Current estimates for such EMRIs extend over a broad range between $10^{-9}$ and $10^{-6}$ per galaxy per year \citep{1997MNRAS.284..318S,2002MNRAS.336..373I,2004CQGra..21S1595G, HA05,2009ApJ...698L..64K}. 
Refinement of these estimates is problematic as there are substantial unknowns about the rates, including uncertainties in the density profiles of galactic nuclei, the populations of compact remnants and more \citep{2007CQGra..24R.113A}. Galactic nuclei are commonly modeled with weak mass segregation, where the stars form mass-dependent cusps near the MBH, $n\propto r^{-\gamma_m}$, with indices $\gamma_H = 1.75$ for the heavy stars and $\gamma_L = 1.5$ for the light stars \citep{1977ApJ...216..883B}.

In this work, we consider several galactic nucleus models, including models of nuclei that experience strong mass-segregation, with indices for the heavy and light populations as shown in Table~\ref{tab:newT}. 
In addition, we study the population of COs captured into a close bound orbit near the MBH following a binary disruption event \citep{Hills1988}. Introducing captured stars might change the overall distribution of stellar populations if the capture rate is sufficiently high \citep{2018ApJ...852...51F}, but mass-segregation effects could potentially smear this effect; here we assume the nuclear cluster is not significantly affected by captured stars. Following a close approach to the MBH, stellar-mass binaries containing a CO component, come sufficiently close to the MBH such that the binary is tidally separated, leaving one object bound to the MBH and the other ejected to infinity at high velocity \citep{Hills1988,Yu_2003}; although the rates of such disruptions could be low, various processes can significantly increase them \cite[e.g.][]{Per07,Ham17}. The stellar object that remains bound to the MBH, is captured into a high eccentricity orbit.  Those stars captured onto orbits with close approaches to the MBH might gradually sink onto the MBH via GW emission, circularize and inspiral to the MBH before stellar orbital perturbations significantly deflect their trajectory. The corresponding event rate could be comparable to that produced by the aforementioned capture of individual BHs from eccentric orbits. In addition, the large periapse distance after capture implies that when tidal separation EMRIs are detected with LISA, they will have eccentricities close to zero \citep{0eccLISA}. Consequently, generation of EMRIs through tidal separation of binaries potentially has a distinct signal. 

In what follows, we adopt an MC approach to investigate the generation of EMRIs through binary tidal separation. Section~\ref{sec:theory} recapitulates relevant results of loss cone theory and dynamical relaxation in galactic nuclei (\ref{sec:tr}), presents our modeling of the distribution functions adopted in the study (\ref{sec:masseg}) and lays out the theoretical framework for tidally split EMRIs 
(\ref{sec:Bsep}). Sec.~\ref{sec:MC} describes the MC simulation. In sec.~\ref{sec:results} we apply the simulation to an MBH in galactic nuclei that experience either weak or strong mass segregation, and detail the results of our investigation, which are then summarized in Sec.~\ref{sec:sum}.

\section{Theoretical framework} \label{sec:theory}
Supermassive black holes, which are thought to reside at the center of most galaxies, act like sinks by removing stars that come sufficiently close to them. This removal can occur in one of two ways, depending on the mass of the MBH and on the properties of the star. Bound COs that are not disrupted by tidal forces of the MBH, will eventually be swallowed whole by the MBH, i.e. find themselves on loss-cone\footnote{The phase-space volume $J < J_{lc}$ is known as the ``loss cone''. We assume a non-rotating MBH, and note that capture of compact objects is more likely to be preceded by scattering into highly eccentric, i.e. nearly zero-energy orbits. We therefore use in our simulations the reasonable value of $J_{lc}=4GM\bh/c$, which is nearly constant over the relevant range of the specific energy.} orbits that take them inside the MBH event horizon. These objects either become direct plunges, or gradually inspiral due to the emission of GWs \citep{HA05}, i.e. become EMRIs. 


\begin{table*}
  \begin{tabular}{|cccccccccc|}
		\hline 
		Star & NSC & $m'$ & $\gamma'$ & $\gamma_{\ast}$ & $a_{norm}$ & $N(a_{norm})$ & $a_c$ & $\Gamma_{ins}$, MC & $\Gamma_{ins}$, lit. \\
		 & model & $(M\ms)$ & & & (pc) & & (mpc) & (Gyr$^{-1}$) & (Gyr$^{-1}$) \\
		\hline
		WD & BW & 0.6 & 1.5 & 1.5 & 2 & $0.1\times N_h$ & 6.3 & 16.6 & - \\
		NS & BW & 1.4 & 1.5 & 1.5 & 2 & $0.01\times N_h$ & 10.6 & 3.4 & - \\
		BH & BW & 10 & 1.75 & 1.5 & 2 & $0.001\times N_h$ & 37.7 & 1.95 & - \\
		WD & HA06 & 0.6 & 1.5 & 2.0 & 0.1 & 2700 			& 3 & 25 & 30$^\dagger$ \\
		NS & HA06 & 1.4 & 1.5 & 2.0 & 0.1 & 374 
		& 4 & 4.5 & 6$^\dagger$ \\
		BH & HA06 & 10 & 2.0 & 2.0 & 0.1 & 1800 
		& 13 & 255 & 250$^\dagger$ \\
		WD & AP16 & 0.6 & 1.3 & 1.9 & 0.1 & 22780 & 0.165 & 21 & 71 $^\ddagger$ \\
		NS & AP16 & 1.4 & 1.3 & 1.9 & 0.1 & 1315 & 0.5 & 4.7 & 11 $^\ddagger$ \\
		BH & AP16 & 10 & 1.5 & 1.9 & 0.1 & 800 & 3.5 & 33 & 92 $^\ddagger$ \\
		BH & AP16 & 30 & 1.9 & 1.9 & 0.1 & 1335 & 8.8 & 308 & 265$^\ddagger$ \\
		\hline
  \end{tabular}
\begin{footnotesize}
\begin{flushleft}
$^\dagger$ \cite{HA06seg}. \\
$^\ddagger$ \cite{Danor}.
\end{flushleft}
\end{footnotesize}

\caption[]{The parameters of the CO populations in the nuclear star clusters, and the predicted rates of GW events. The MBH mass is taken to be $M\bh=4\times 10^6$. AP16 refers to the  right panel of Fig.~1 in \cite{Danor}. In the BW cusp, the rate for WDs is highest, but BH EMRIs dominate the rates for both the weakly and strongly mass-segregated regimes, and the overall rates are significantly enhanced overall, compared with the simple BW model. It is also important to mention the hierarchy of the rates is not necessarily the same as the cosmic {\emph observable} rates that LISA will observe, since NSs and BHs are more massive than WDs and can be observed at larger distances. Note that the rates are substantially higher for nuclear clusters which experience strong mass segregation.} 
\label{tab:newT}
\end{table*}

\subsection{Two-body relaxation} \label{sec:tr}
The stellar orbits are defined by a specific angular momentum $J$ and relative specific energy $\epsilon$. We work under the assumption that stars distribute isotropically around the MBH.

In a spherical nuclear cluster, the number of stars with angular momentum small enough to satisfy $J\lesssim J_{lc}$ would ordinarily be small; furthermore, the loss cone at a given energy would be emptied in one orbital period and no further stars would be lost to the MBH.
In realistic systems, however, there is a continued supply of stars into the loss cone. 
An often-discussed mechanism for loss-cone re-population is 
gravitational encounters between stars, ``two-body relaxation'', which cause their energy and angular momentum to gradually evolve until they enter the loss cone.
For an EMRI to occur, in the standard picture, two-body relaxation has to bring a compact remnant onto an orbit with such a small pericenter distance that dissipation of energy by emission of GWs becomes significant. 

\paragraph*{}Diffusion in $\epsilon$-space occurs on the relaxation timescale $t_r\sim \epsilon/\dot{\epsilon}$, whereas diffusion in $J$-space occurs on the angular momentum relaxation timescale
\be t_J\sim \frac{J}{\dot{J}}\sim \left[ \frac{J}{J_m(\epsilon)}\right]^2t_r, \label{eq:tj} \ee 
where $J_m(\epsilon)$ is the maximal (circular orbit) angular momentum for specific energy $\epsilon$. 

Diffusion in $J$-space is much more efficient than in $\epsilon$-space \citep{2017ARA&A..55...17A}. We therefore follow \cite{2019MNRAS.485.2125B}, who computed the relaxation time associated with angular momentum changes rather than the typically adopted one. 
Following \cite{HA05}, $t_r$ is defined as the inverse of the orbit-averaged, Fokker-Planck diffusion coefficient. \cite{2019MNRAS.485.2125B} obtained
\be t_r(a)=\frac{3\sqrt{2}\pi^2}{32C_{\gamma_{\ast}}}\left( \frac{GM\bh}{a}\right) ^{3/2}\frac{a^{\gamma_{\ast}}}{G^2m_{\ast}^2N_0\ln{\Lambda}}, \label{eq:tr} \ee 
where $a$ is the orbital semimajor axis. Eq.~(\ref{eq:tr}) has been obtained under the assumption that the number density of the stellar objects inducing angular momentum relaxation scales as $n_{\ast}(a)\propto a^{-\gamma_{\ast}},\ \gamma_{\ast} > 0.5$. 
$N_0$ is the normalizing constant to the number of stars within a given $a$, i.e. $N(a)=N_0a^{3-\gamma_{\ast}}$. $C_{\gamma_{\ast}}$ is a dimensionless constant of the order of unity for relevant $\gamma_{\ast}$ values, whose value as a function of $\gamma_{\ast}$ is plotted in Fig.~A1 of \cite{2019MNRAS.485.2125B}. $m_{\ast}$ represents the typical mass of stars that dominate the two-body relaxation, and 
$\ln{\Lambda}$ is the Coulomb logarithm, which can be approximated as $\ln{\Lambda} = \ln{[M\bh/(2m_{\ast})]}$ within the MBH sphere of influence (following \citealp{PhysRevD.84.024032}).

\paragraph*{} Eq.~(\ref{eq:tr}) showcases that even within the most studied and best understood standard picture of EMRIs, there are substantial unknowns about the rates; expressed by the relaxation time, including uncertainties in the density profiles of galactic nuclei, the populations of compact remnants, number density of MBHs and more \citep{2007CQGra..24R.113A}. In the section that follows, we describe three different models of galactic nuclei.

\subsection{Mass segregation} \label{sec:masseg} 
Prevailing theory and observations suggest that galactic nuclei host stars with a wide range of masses, and that mass segregation processes likely play an important role in shaping the density profiles of nuclear clusters \cite{2011CQGra..28i4017A}. Mass-segregation processes proceed through two-body relaxation where stars of different masses encounter each other. This gives rise to the redistribution of the  orbital energies among stellar objects, such that the more massive objects migrate closer to the MBH, while the lighter ones drift to larger radii. Mass segregation plays a key role in EMRI analyses, since it increases the density of the more massive COs within the region where they could inspiral and become LISA sources.

The quasi-steady state of a mass segregated cusp near an MBH can generally be described by a power-law density distribution $n_*(r)\propto r^{-\gamma_*}$ with a slope that is steeper for heavier objects. Previous studies primarily investigated cases of weak mass segregation, hereafter referred to as the BW-cusp, where the heaviest objects form a power-law density cusp with $\gamma'=7/4$ around the MBH, while less massive species arrange themselves into a shallower profile, with $\gamma_{\ast}=1.5$ \citep{1977ApJ...216..883B}. 

\begin{figure}
	\includegraphics[width=\columnwidth,clip,trim={0.45cm 0.4cm 1.05cm 0.83cm}]{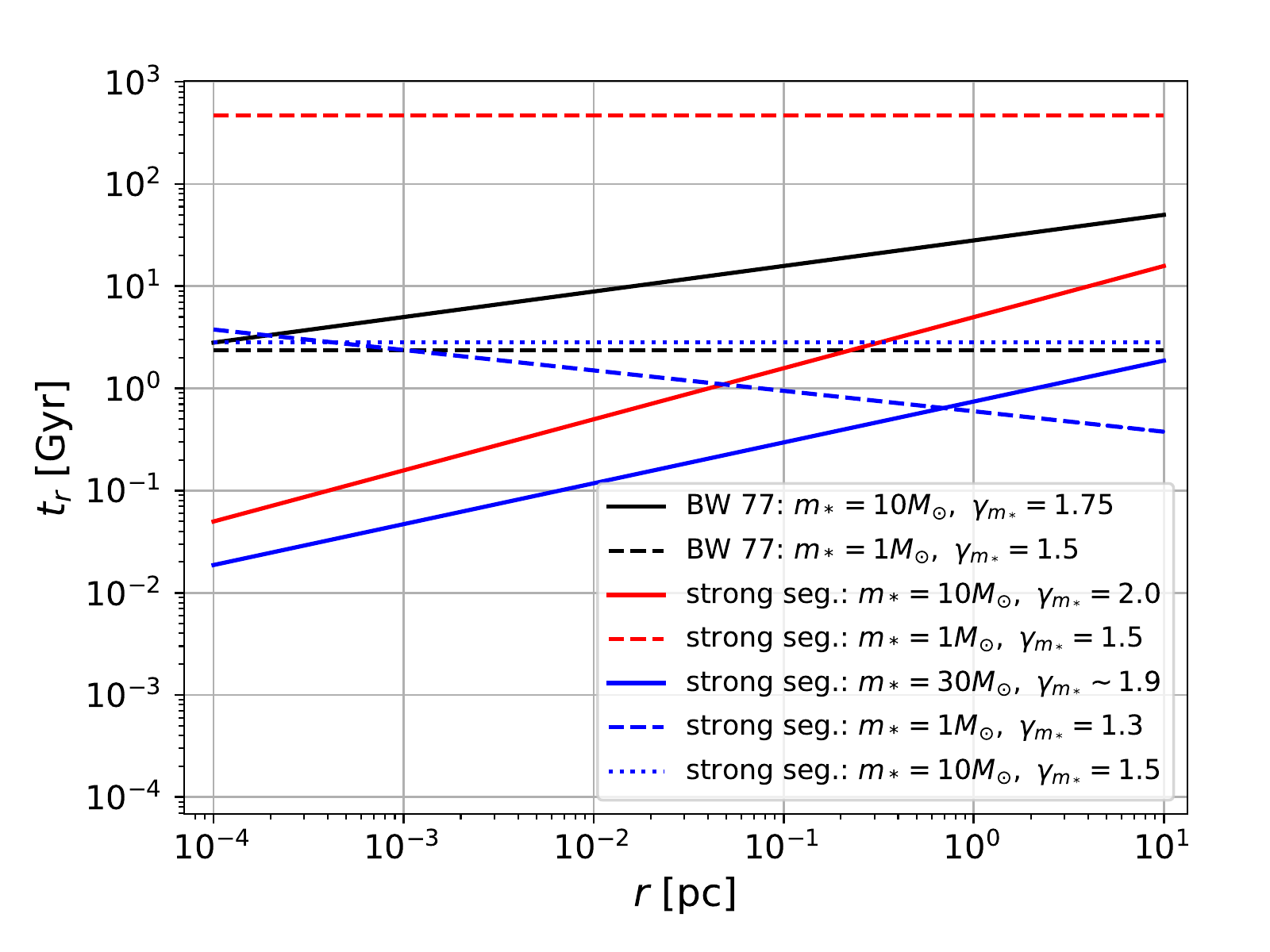}
    \caption{Relaxation timescale [$t_r$, computed via Eq.~(\ref{eq:tr})], as a function of the semimajor axis of stars orbiting a $4\times 10^6 M\ms$ MBH. The relaxation timescale is evaluated assuming that relaxation effects are induced either by a steep cusp of relatively heavy stellar BHs (solid lines) or by a shallower cusp populated by $1 M\ms$ stars (dashed lines); the three NSC models and respective quantities adopted for the computation of $t_r$ are detailed in Table~\ref{tab:newT}.}
    \label{fig:tr}
\end{figure}

In the strong mass-segregation regime, the massive objects can achieve an even steeper density profile compared with the weakly-segregated case \citep{2009ApJ...698L..64K,Pre10,Danor}. Stronger mass segregation increases the concentration of the more massive COs within the region where they could become detectable EMRIs. Moreover, rather than being dominated by low-mass stars, relaxation around the MBH, in particular in the closest regions near the MBH, might be dominated by the heaviest COs in strong segregated cusps, and the timescale for relaxation can become shorter in these regions. Therefore, EMRI rates can be significantly enhanced. Although the effects of the various mass-segregation regimes on the EMRI rates have been studied before, the impact on the EMRI eccentricity distribution, have not been explored, nor the contribution of captured stars in these cases.

The above considerations motivate us to consider three different models of nuclear stellar cluster (also depicted in Table~\ref{tab:newT}); a BW cusp in which stellar-mass BHs are relatively common such that relaxation is dominated by lighter objects and $t_r\simeq 2$~Gyr (see Fig.~\ref{fig:tr}), an NSC that contains only a small fraction of $10 M\ms$ BHs which sink to the center (by dynamical friction) and form a much steeper cusp, and a model that includes two populations of $10$ and $30 M\ms$ BHs (and the latter dominates relaxation). The second and third models correspond to the weakly and strongly segregated cases as obtained by \cite{HA06seg} and \cite{Danor} respectively, and described in their Figure~1.

\paragraph*{} There are other possible modifications of the standard description now being included in EMRI analyses, such as resonant relaxation. 
In what follows, however, we neglect this effect as it was recently shown to have a negligible effect in the space parameter of interest \citep{2016ApJ...820..129B}.

\subsection{Captured stars following binary disruption events and EMRIs} \label{sec:Bsep} 
Binaries containing at least one BH that pass close to the MBH, get tidally separated; one member of the binary remains bound to the MBH on a highly eccentric orbit, while the other is ejected as a hyper-velocity star with velocities of hundreds km\,s$^{-1}$ \citep{Hills1988,Yu_2003}. The rates of such binary disruptions could be low, however various processes such as perturbations by massive perturbers and nuclear spiral arms could significantly increase these rates \citep{Per07,Ham17}. 

The key point about this process is that it can potentially increase the number of highly eccentric stars very close to the MBH, such that their orbits are close to the loss cone, and could significantly evolve due to GW emission.  

In order to study the orbit of a captured star, let us consider a binary that is disrupted at a distance $r_{t,bin}=a_{bin}(M\bh /m_{bin})^{1/3}$. Here $a_{bin}$ is the semimajor axis of the binary pre-separation and $m_{bin}$ is the total mass of its components. The point of tidal disruption becomes the periapsis of the new orbit.

The semimajor axis of the captured object following its capture is \citep{2013degn.book.....M}
\be a_{capt}\approx \left( \frac{M\bh}{m_{bin}}\right) ^{2/3}a_{bin}. \ee
This relation maps the semimajor axis distribution of the infalling binaries to that of the captured stars: the tighter (closer) the binaries, the more tightly bound are the captured stars.

Hence, following the binary disruption of a typical stellar binary, the eccentricity of the captured object is \citep{1989AJ.....97..222H}
\be e_{capt}=1-\left( \frac{M\bh/4\times 10^6\,M_{\odot}}{m_{bin}/20\,M_{\odot}}\right) ^{-1/3}\sim 0.98. \ee 
Most interestingly, \cite{0eccLISA} pointed out that the GW-inspiral time of compact objects captured through this process very close to the MBH, could be sufficiently short, such that they are not perturbed by other stars, and the GW-inspiral leads to the circularization of their orbits and eventually to their inspiral to the MBH. Thereby, these tidally-split EMRIs should be very close to circular when they have shrunk into LISA band. Although stars which migrated to similar close and eccentric orbits through two-body relaxation could also produce such low-eccentricity GW-EMRIs, the tidal binary-disruption mechanism can significantly enhance the rate of such low-eccentricity EMRIs, far beyond the case of a relaxed nuclear cluster without any contribution from binary-disruption processes.
However, \cite{0eccLISA} considered nuclear clusters without including mass-segregation, and considered only approximately the effects of two-body perturbations on the inspiraling COs. In that case the direct capture of COs, and in particular BHs, could lead to a significant increase of GW sources, however, as we show here, when mass-segregation is accounted for, BHs are far more abundant close to the MBH, and the relative contribution from captured sources is far lower. Moreover, two-body perturbations are more effective in changing the orbits of captured-stars than suggested by \cite{0eccLISA} leading to higher eccentricities than estimated by them. Nevertheless, for most cases and stellar species considered here, the overall eccentricity distribution of EMRIs originating from captured stars is still centered around lower eccentricities than the general EMRIs population arising from non-captured stars in the nuclear cluster. 

\section{Monte-Carlo calculations} \label{sec:MC} 
To probe the orbital evolution of compact objects in galactic nuclei that harbor an MBH, we adopt an MC approach first suggested and used by \cite{1995ApJ...445L...7H} and later-on refined by \cite{HA05} \cite[see also][]{1978ApJ...225..603S,2016ApJ...820..129B}. The simulations follow the evolution of orbits of test-particles subject to GW emission \citep{PhysRev.136.B1224} and two-body relaxation introduced by random perturbations of the angular momentum according to pre-computed ``diffusion coefficients'', as shown below (diffusion in energy is neglected since the stars are in highly eccentric orbits, and the angular momentum diffusion is far more important in producing EMRIs).

Throughout this paper, we focus on galactic nuclei similar to the Galactic center of our Galaxy; we assume a static cusp profile, and a non-spinning\footnote{MC simulations with Kerr metric orbits show that the overall results hold \citep{HA05}.}, $M\bh= 4\times 10^6M\ms$ MBH.

The simulation follows a star on a relativistic orbit, described by \citep{HA05}
\be \epsilon^2_{GR}=\frac{(q-2-2e)(q-2+2e)}{q(q-3-e^2)}, \ee
\be J^2=\frac{q^2}{q-3-e^2}\left( \frac{GM\bh}{c}\right) ^2 \ee
where $q=a(1-e^2)c^2/(GM\bh)$ \citep{1994PhRvD..50.3816C}.

Assuming a static cusp profile, the orbits of the test-particles are subject to GW emission, Eqs.~(\ref{eq:dJGW}--\ref{eq:dEGWg}), and two-body relaxation introduced by random perturbations of the angular momentum according to pre-computed ``diffusion coefficients'' (diffusion in energy is neglected).
Only the ``diffusive regime'', where stars slowly diffuse in $J$-space and the loss cone remains nearly empty at all times, is relevant for inspiral. 
The step in $J$-space per orbit is therefore the sum of three terms:
\be \delta J(\epsilon,J)=\Delta_1J_{scat}(\epsilon,J)+\chi\Delta_2J_{scat}(\epsilon,J)-\Delta J_{GW}(\epsilon,J). \label{eq:dJ} \ee
The first and second terms represent two-body scattering, with a drift term $\Delta_1J_{scat}=\langle\Delta J\rangle P=J^2_mP(\epsilon)/(2t_rJ)$, $P$ is the orbital period and $\Delta_2J_{scat}=[\langle(\Delta J)^2\rangle P]^{1/2}=[P(\epsilon)/t_r]^{1/2}J_m(\epsilon)$. The random variable $\chi$ takes the values $\pm 1$ with equal probabilities. The third term is the deterministic angular momentum loss due to GW emission, defined by \cite{PhysRev.136.B1224}
\be \Delta J_{GW}=-\dfrac{16\pi}{5}g(e)\frac{Gm'}{c}\left( \frac{r_p}{r_s}\right) ^{-2}, \label{eq:dJGW} \ee
where $r_p$ is the periapse distance, $r_s\equiv 2GM\bh/c^2$ and
\be g(e)=\frac{1+(7/8)e^2}{(1+e)^2}. \label{eq:dJGWg} \ee
The energy step per orbit is: 
\be \Delta E=\Delta E_{GW}=\dfrac{8\pi}{5\sqrt{2}}f(e)\frac{m'c^2}{M\bh}\left( \frac{r_p}{r_s}\right) ^{-7/2}, \label{eq:dEGW} \ee
\be f(e)=\frac{1+(73/24)e^2+(37/96)e^4}{(1+e)^{7/2}}. \label{eq:dEGWg} \ee
Note, however, that HA05 introduced a drift-term with an erroneous,  negative sign\footnote{We would like to thank Wenbin Wu, for pointing out this mistake.}, which we corrected in our calculation; the erroneous calculation does make a non-negligible effect on the results as we discuss below.

\cite{HA05} parametrized the relative importance of dissipation and scattering by the ratio of the inspiral time $t_0=2\epsilon_0P_0/\Delta E$, defined as the time it takes the initial energy $\epsilon_0$ to grow formally to infinity, to the angular momentum relaxation timescale (Eq.~\ref{eq:tj}),
\be s(J,a)\equiv \frac{t_0(J,a)}{t_J(J,a)}. \ee
Usually, it is assumed that stars spiral in without further perturbations once $s=1$ (e.g., \citealp{2001CQGra..18.4033F}). However, since large scatterings continue to redistribute the orbital parameters at that stage, such assumption might lead to an overestimate of the total event rate, as stars that actually fall into the MBH are erroneously counted as EMRIs. Moreover, we find that the final eccentricity of the GW sources as they enter into the LISA band is still affected by (albeit more rare) perturbations even for $s=1$.  In our simulations we adopt the value required by \cite{HA05} of said ratio, $s_{crit}=10^{-3}$, in order to ensure the EMRI will be successful, and that the correct eccentricity will be registered. 
Indeed, we note that the premature neglect of the effects of scattering may also explain the underestimate of detectable eccentricities of capture-born EMRIs in \cite{0eccLISA}, though their criterion for inspiral was not clearly specified, as we discuss below.

\subsection{Nuclear star clusters}
In order to study the orbital evolution of COs drawn from the background population of stars in the nuclear cluster near the MBH, we considered a grid of semi-major axes for the initial orbit of the stars, and sampled the initial eccentricities of the stars from a thermal eccentricity distribution. For each semi-major axis, we ran $\mathcal{O}(10^4)$ MC realizations with a different set of randomly chosen perturbation and followed the evolution of the stellar trajectories through phase space.

We stopped the simulations when the orbital period fell below the longest period detectable by LISA, $P_L= 10^3\,$s\footnote{As EMRIs are expected to be weak sources, the overwhelming majority of them will be detected at gravitational wave frequencies greater than $2\times 10^{-3}\,$Hz, i.e. beyond the range of the unresolved Galactic double white dwarf foreground \citep{2001A&A...375..890N,2003MNRAS.346.1197F,2004PhRvD..69h2005B}. The orbital frequencies of detectable EMRIs should therefore be greater than $10^{-3}\,$Hz.}, and recorded the eccentricity at that point and the fraction of stars that avoid falling in the MBH, $S(a_0)$, thereby obtaining the distribution function $W (e ; a_0)$ and the critical semimajor axis $a_c$, under which the majority (>0.5) of COs will eventually become EMRIs. The integrated distribution over all cusp stars, $W(e)$, is then obtained by adding together all the distributions, weighted by $n(a_0)S(a_0)da_0$, where $n(a)$ is the radial integrand of $N(a)$.

Given $S(a_0)$ and $a_c$ we estimated the EMRI rates using 
\be \Gamma = f_s\int_0^{\infty}{\frac{n(a) S(a)da}{\ln{(J_m/J_{L})}t_r(a)}}\simeq f_s\int_0^{a_c}{\frac{n(a)da}{\ln{(J_m/J_{lc})}t_r(a)}}, \label{eq:rate} \ee
where roughly $S(a_c)\sim 0.5$, and $f_s$ is the number fraction of stars of type $s$ \citep{HA05}. 

\subsection{Stars captured a binary tidal disruption event} 
In order to simulate EMRIs arising from captured stars, we initialized our simulations with the expected parameters for captured objects after separation. We note that the distribution of semimajor axes after capture is debated\footnote{Mainly since it depends on the distribution of semimajor axes of binaries containing BHs \citep{0eccLISA}.}.
Here we consider a uniform in log distribution for the binary separation, $a_{bin}$, between $0.05$ and $30$ AU\footnote{Binaries of $10M\ms$ BHs that are tighter than $a_{bin}=0.05$ AU will very likely merge by gravitational radiation before being separated by the MBH. On the other hand, very wide binaries (more than a few AU) could end up after separation with semimajor axes so large that perturbations during a single dynamical time drop them into plunge orbits, making them undetectable with LISA.}, which is consistent with \cite{2012Sci...337..444S,2012ApJ...757...27A}'s distributions. The initial eccentricities were taken to be high with $e_{init}\approx 0.98$. 

\subsection {Code verification}
In order to benchmark our calculations, we first attempted to reproduce HA05 results. Since they used an erroneous drift-term in their calculation, we first used the same type of erroneous calculations, and then reran our MC with the corrected term. This also allowed us to quantify the effect of the error. Though far-less important, we also note that HA05 used a lower mass MBH with $3\times 10^6\,{\rm M}_{\odot}$ (consistent with the best mass-estimate of the MBH at the time), which we therefore adapted in our calculations when making the direct comparison. 

\begin{figure}
\includegraphics[width=\columnwidth,clip,trim={0.6cm 0.3cm 1.3cm 1.2cm},clip]{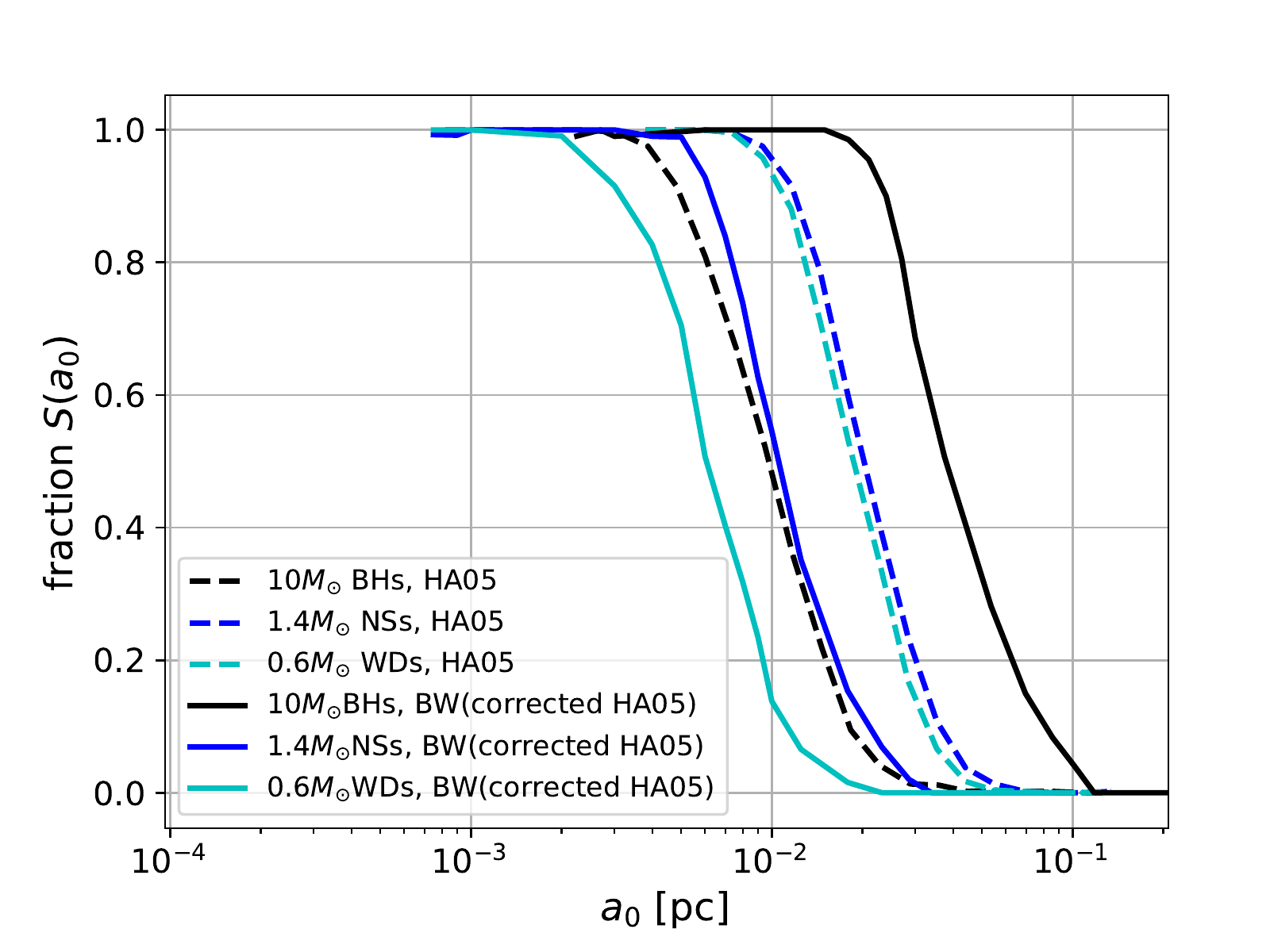}
\caption{The fraction of inspiral events vs. the initial semimajor axes for a BW cusp, before and after correction for the drift-term used in HA05. All calculations refer to a nuclear cluster around an MBH of $M\bh = 4\times 10^6M\ms$. The MC simulations performed with identical characteristics as in HA05 agree with their results. However, as can be seen, using the correct drift-term changes the results.} 
\label{fig:sa0}
\end{figure}

The total number of stars within $r_h=2$\,pc was assumed to be $N_h= 2M\bh/M\ms$, with different number fractions for the respective species. See Table~\ref{tab:newT} for the assumed parameters of the stellar populations. The stars were distributed according to BW power-law distribution with $\gamma_{BH}=1.75,\ \gamma_{rest}=1.5$ \citep{1977ApJ...216..883B}. Fig.~\ref{fig:sa0} shows the normalized inspiral probability function $S_s (a_0)$, where $s$ stands for WD, NS, and BH. 
The lines corresponding to HA05 in Figure \ref{fig:sa0}, as well as our estimates for the inspiral rates, presented in Table~\ref{tab:newT}, are highly compatible with the results obtained by HA05, providing an excellent verification for our calculations. 

\begin{figure}
\includegraphics[width=\columnwidth,clip,,trim={0.55cm 0.2cm 1.2cm 1.2cm}]{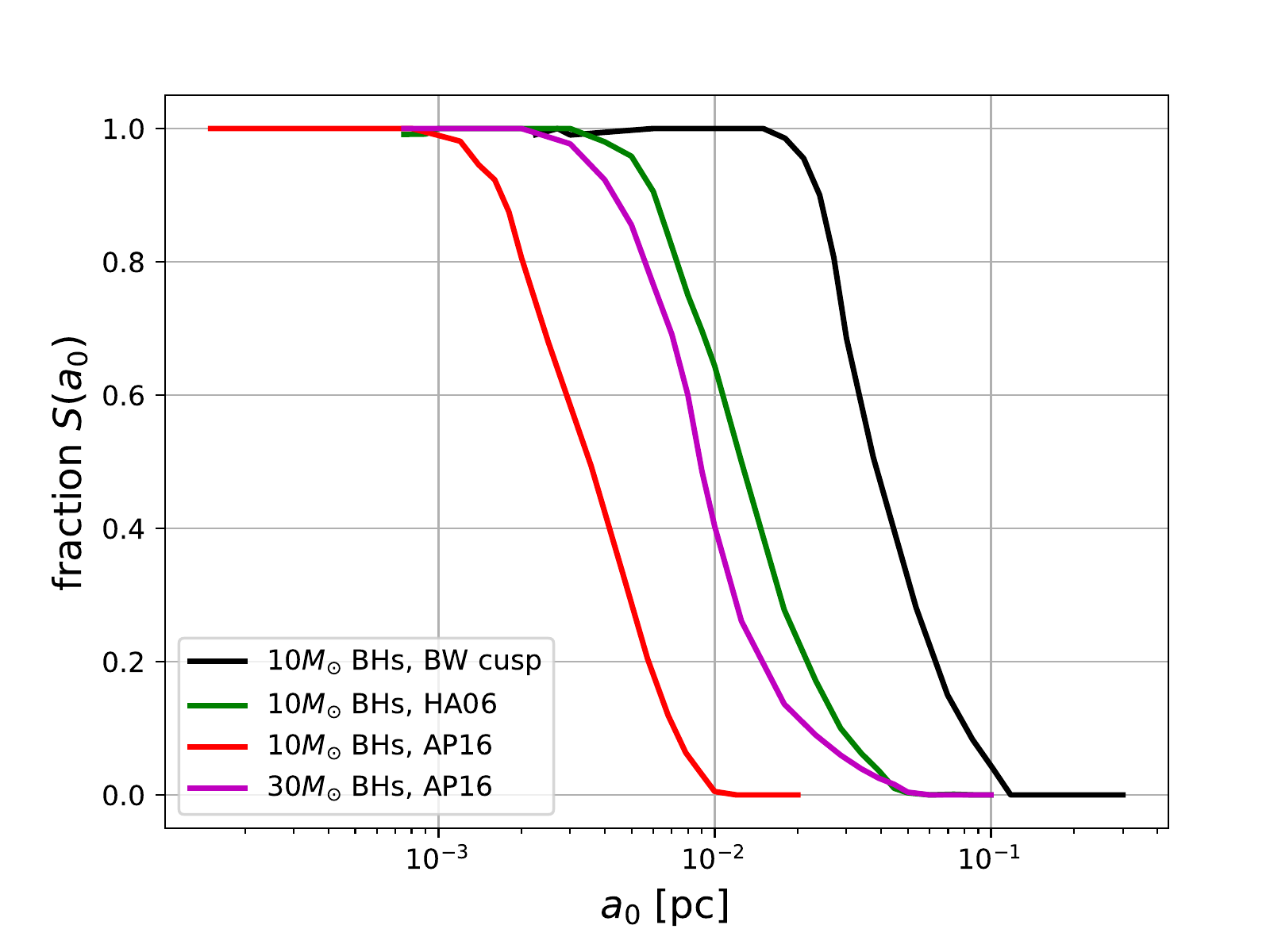}
\caption{The fraction of inspiral events of stellar BHs vs. the initial semimajor axis, for an MBH of $M\bh = 4\times 10^6M\ms$, comparing both weakly segregated and strongly segregated clusters (see Table~\ref{tab:newT}). Compared with the weakly segregated BW case, the critical semimajor axes for stellar BHs, and in particular more massive BHs in nuclei that experience strong mass segregation, are smaller than the ones obtained for shallower density cusps because of the smaller relaxation time and higher central concentration due to stronger mass segregation.
} \label{fig:sa1}
\end{figure}

\section{Inspiral rates and the distribution of the orbital parameters} \label{sec:results}

\subsection{Comparison between weakly and strongly mass segregated clusters}
Qualitatively, the inspiral rate depends on the number of COs inside $a_c$, and so mass segregation is expected to play an important role in enhancing the event rate by leading to a centrally concentrated distribution of BHs. This has been confirmed quantitatively in several past studies \citep{HA06seg,Danor}.

Figure \ref{fig:sa1} shows the normalized inspiral probability function $S_w (a_0)$, where $w$ stands for $10M\ms$ BH in cluster model no. 1, $10M\ms$ BH in model no. 2 and $30M\ms$ BH in model no. 2. As expected, the steeper the cusp, the higher the rates of inspirals of the more massive COs. As mentioned above, we also show the results from clusters modeled following \cite{HA06seg,Danor}. The rates for the BW cusp (done with the correct drift-term) give rise to higher rates of BH EMRIs, but lower rates of NS and WD EMRIs compared with HA05 (see table \ref{tab:newT}).

We also applied the MC simulation to nuclei that experience weak and strong mass segregation; 
to that end we adopted two cluster models from \cite{HA06seg} and \cite{Danor}, described in their Fig.~1. The properties of the two cluster models, as well as the calculated rates of GW events, are summarized in Table~\ref{tab:newT}.  
The relaxation time $t_r$ as a function of $a$ is displayed in Fig.~\ref{fig:tr}; relaxation effects in central areas of the clusters are induced by a steep cusp of heavier stellar BHs. As can be seen, strongly mass-segregated clusters give rise to far higher GW EMRI rates, and the inclusion of even more massive BHs, bias the EMRIs towards inspirals of the more massive BHs, while quenching the inspirals of less massive ones. We find this effect, suggested in \cite{Danor} to be much more pronounced than found there. Note that \cite{Danor} only considered a simplified approach for assessing the EMRI rates, and did not follow the detailed evolution using the MC approach we use here. In addition, they made use of somewhat different assumptions on the BH, NS and WD fractions, and get effective somewhat different relaxation times. These can explain the slight difference in the expected rates.  

\begin{figure*}[h]
\includegraphics[scale=0.74,trim={1.6cm 0.7cm 0.5cm 0.5cm}]{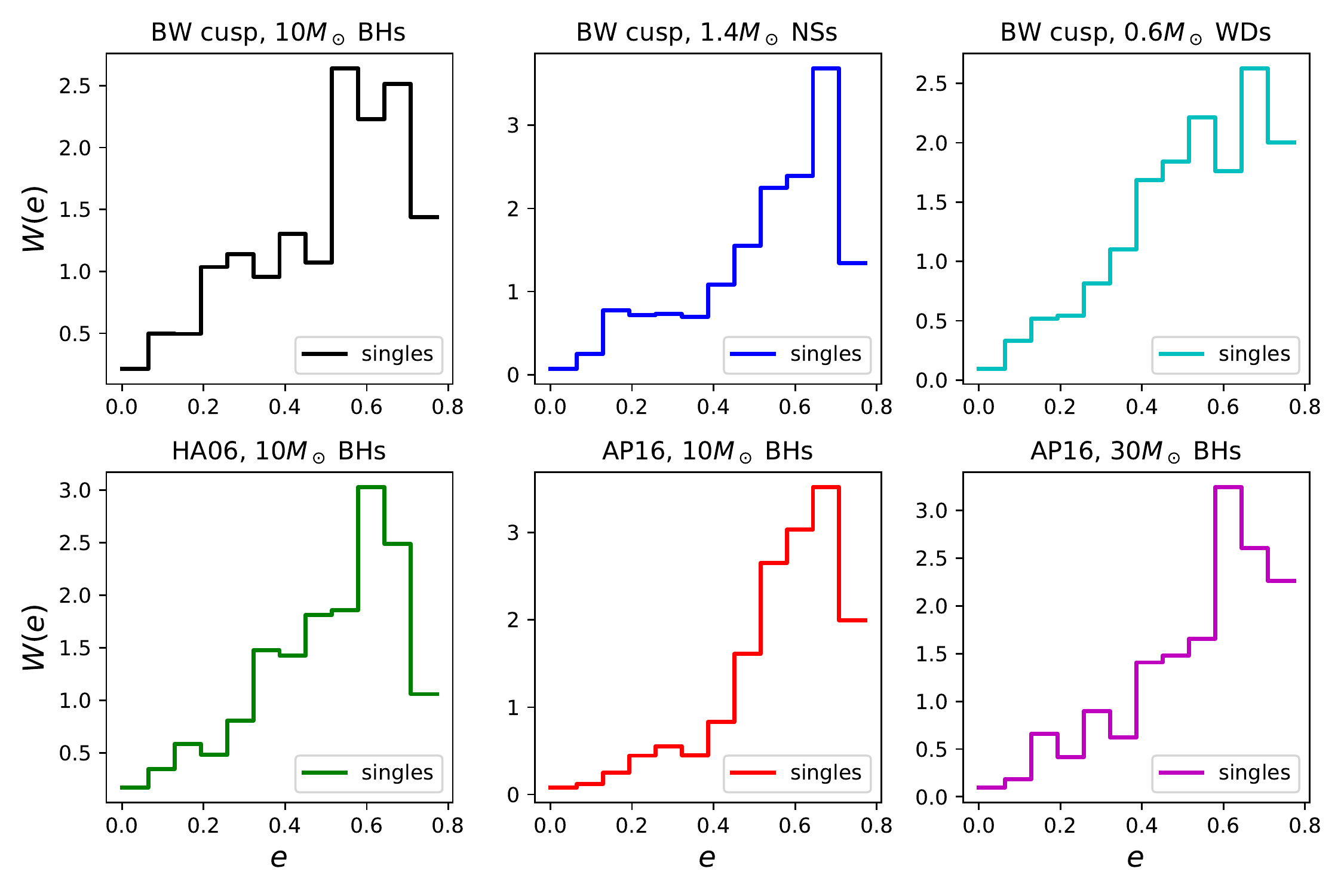}
\caption{Detectable eccentricity distributions (of inspirals at gravitational wave frequencies of $2\times 10^{-3}\,$Hz) of EMRIs in a relaxed BW cusp (top) and relaxed cusps that experience strong mass segregation (bottom).} 
\label{fig:ecc_sng}
\end{figure*}

Besides the rates, we show the distribution function of EMRIs eccentricities in the LISA band in Fig.~\ref{fig:ecc_sng}. As found in earlier studies, eccentricities are skewed towards high values. We find this is still the case for the corrected calculation of the BW-cusp case (shown for BHs, NSs and WDs) as well as the strong-mass segregation case (shown only for BHs). Although observable differences can be seen between the weak and strong mass-segregation regimes, the overall structures are similar. As we discuss in the next section, captured stars can give rise to lower eccentricities.

\subsection{Binary tidal separation and zero/low-eccentricity LISA events} \label{sec:methBsep}

\begin{figure*}
\includegraphics[scale=0.74,trim={1.6cm 0.7cm 0.5cm 0.5cm}]{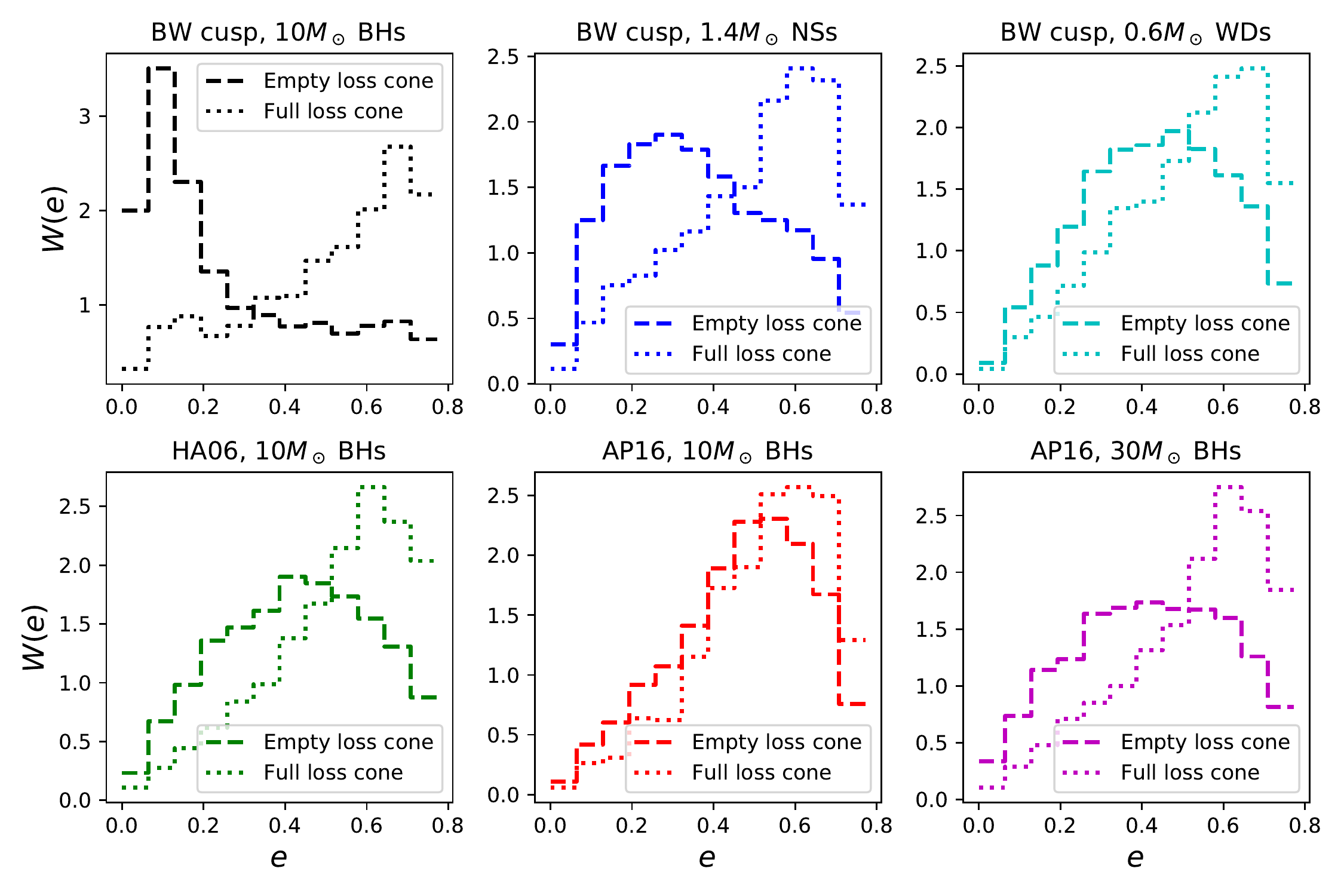}
    \caption{The eccentricity distribution of EMRIs, at gravitational wave frequencies of $2\times 10^{-3}\,$Hz, in a BW cusp (top) and cusps that experience strong mass segregation (bottom), originating from post binary-disruption captured stars, assuming $a_{bin}\propto\ $loguniform, between a few 0.01 to few AU (i.e, binaries that would not merge by gravitational radiation before being separated by the MBH, as well as would not be captured in orbits with semimajor axes larger than $a_{crit}$). We consider both the empty loss-cone regime (dashed), where tidal disruptions directly maps the binary log-uniform distribution into a log-uniform distribution of orbits around the MBH; and the full loss-cone regime (dotted), where wider binaries have a higher disruption probability, and the distribution of the orbits of captured stars is uniform \citep[see][]{Per10}. 
}    \label{fig:ecc_bin}
\end{figure*}

\begin{figure*}
\includegraphics[scale=0.74,trim={1.6cm 0.7cm 0.5cm 0.5cm}]{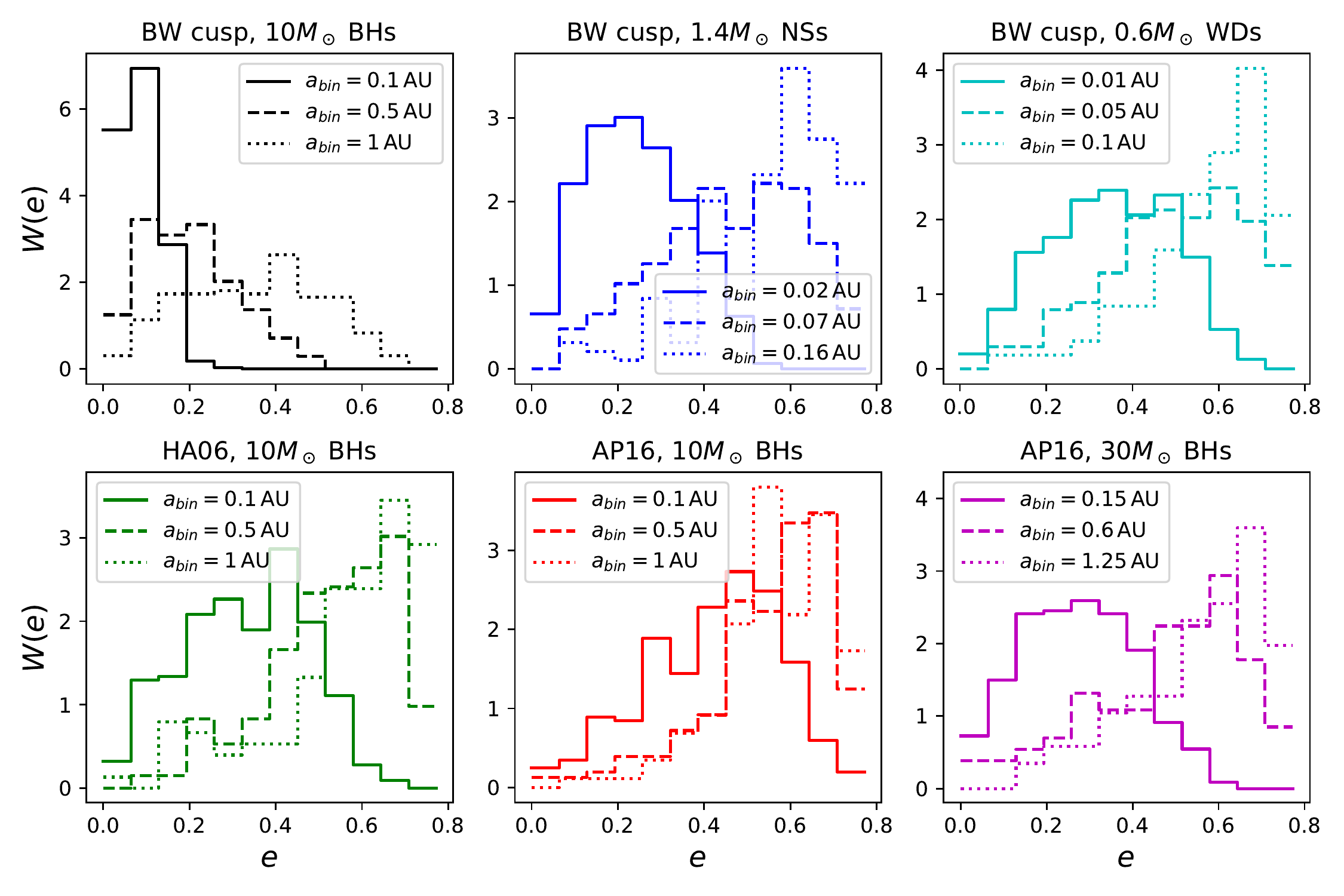}
    \caption{The eccentricity distribution of EMRIs at gravitational wave frequencies of $2\times 10^{-3}\,$Hz in a BW cusp (top) and cusps that experience strong mass segregation (bottom), originating from post binary-disruption captured stars with specific $a_{bin}$ as ascribed in legend, showing that larger capture distances arising from larger binary separations eventually give rise to higher inspiral eccentricities.}    \label{fig:6}
\end{figure*}

We ran the simulation with initial conditions similar to the expected parameters of captured BHs after binary separation. We consider both the empty loss-cone regime, where tidal disruptions directly maps the binary log-uniform distribution into a log-uniform distribution of orbits around the MBH; and the full loss-cone regime, where wider binaries have a higher disruption probability, leading to a stronger dependence on the binary semi-major axis, and thereby the distribution of the orbits of captured stars has a uniform distribution \citep[see][] {Per10}. The initial eccentricity of captured stars is taken to be high, as discussed above. Fig.~\ref{fig:ecc_bin} shows the eccentricity distribution of EMRIs as they enter the LISA band, where both the distributions from the empty (dashed) and full (dotted) loss-cone regimes are shown.  Fig.~\ref{fig:ecc_bin}  shows the distribution of observable eccentricities of EMRIs originating from such post tidal-disruption captured COs, which, as can be seen,  is skewed towards low values; orbits of captured BHs EMRIs typically have $e\sim 0.1$ in the LISA sensitivity band. Therefore, though not as distinct as past studies estimated \citep{0eccLISA}, the signal of such capture-originating EMRIs is still quite distinct, in comparison with EMRIs from relaxed cluster stars seen in Fig. \ref{fig:ecc_sng}.   

Note that the binaries which lead to a capture close to the MBH, and that can later become EMRIs, need to be relatively tight. Such binaries are relatively hard binaries in the regions where most binaries are scattered onto the MBH (which are typically the empty to full loss-cone transition regions). While the transition region for single stars is typically close to the radius of influence of the MBH, the larger binary separation gives rise to a larger typical transition radius from the MBH, where such tight binaries are hard (the velocity dispersion is smaller), and relatively few are affected by encounters with other stars. The effect of softening is therefore generally small for the binaries contribution to the EMRI populations, and especially stellar BHs. Such effects could, however,  be more important for lower mass NS or WD binaries, which are softer, and become wider due to stellar perturbations, and even evaporate; see discussion in e.g. \cite{Per07} and \cite{Per09}. In order to show the different contribution from smaller vs. larger binary separations, Fig.~\ref{fig:6} shows the eccentricity distribution of EMRIs arising from binaries of different separations: 0.1, 0.5 and 1 AU. The effect of binary separation can be seen in the different results for smaller vs. larger separations. Wider binaries give rise to captures at larger distances from the MBH; the peri-center approach of such stars have larger random walk steps, and are generally less affected by the GW circularization before their final inspiral. Consequently, on average, the larger the capture distance, the less circular would be the final EMRI, consistent with the results shown, and the observed general trend for the full-loss cone (in Fig. \ref{fig:ecc_bin}) to show higher eccentricities (as the binary disruption rate is biased towards larger separations), compared with the empty loss-cone.

Whether the contribution of EMRIs from captured stars can significantly affect the observed distribution of EMRIs in LISA depend on the unknown capture rates, and the cusp structure. In cases where the binary disruption rates are enhanced \cite[e.g.][]{Per07,Mad+09,Ham17}, and the cusp is not strongly segregated, the fractional contribution from captured stars is larger, while for strongly mass-segregated regimes and non-enhanced disruption rates the contribution from relaxed cusp stars is far larger, and is likely to smear any contribution from low-eccentricity EMRI sources from captured stars. Given a specific model for the binary disruption rate, one can weight the contribution from cusp stars and captured stars to obtain the overall distribution using our results.

\section{Summary and discussion} \label{sec:sum} 
Extreme mass ratio GW inspirals of objects around MBHs in galactic nuclei are expected to provide information on the orbital parameters of these objects and shed light on dynamical relaxation and other dynamical processes in galactic nuclei. In this paper we studied the rates and properties of such EMRIs. We considered different environmental properties of the nuclear clusters, and explored both weakly segregated and strongly mass-segregated nuclear clusters (as well as cases with lower and higher masses of stellar BHs in these environments). In addition we studied the potential contribution of compact objects captured into orbits near the MBH following a binary disruption, which can be initially ejected at highly eccentric and relatively short period orbits.

We followed the orbits of stars near an MBH through a Monte-Carlo diffusion approach, identifying when such stars plunged to the MBH, inspiraled to the MBH to produce EMRIs, or survived in the nuclear cusp. We modelled the evolution of the stellar orbits, where we considered the effects of random scattering by stars and orbit dissipation through GW-emission, and recorded the eccentricities of inspiraling objects when they entered the LISA GW-band.  

Consistent with previous results, but now supported by the more detailed diffusion models, We find that strong mass-segregation of nuclear clusters generally leads to significantly higher rates of EMRIs, especially in respect to the most massive stellar BHs in the clusters, compared with weakly segregated clusters.  In addition, provide for the first time the eccentricity distribution of EMRIs in the weak and strong mass segregation regime [and correct previous studies \cite{HA05}, which used an erroneous drift term].  We find the EMRIs eccentricity distribution differs between weakly and strongly segregated clusters. We also considered the contribution of post binary tidal-disruption captured compact objects, and find that although they rarely produce close to zero-eccentricity EMRIs as suggested by previous studies \citep{0eccLISA}, their eccentricity distribution is indeed biased towards significantly lower eccentricities than that of stars originating in the relaxed cluster.     

Taken together we demonstrate and predict the potential of the future space-born GW detectors in both detecting EMRIs, and make use of their statistical properties to explore and characterize the dynamical processes occurring near MBHs in galactic nuclei.  

\section*{Data availability}
The data that support the findings of this study are available from the corresponding author upon reasonable request.


\bibliographystyle{mnras}
\bibliography{EMRIs0e}


\bsp
\label{lastpage}
\end{document}